\documentclass{aa}
\usepackage{graphicx}
\def\der{{\rm d}}
\begin{document}
\thesaurus{02         
              (12.03.2;   
               12.03.4)}  
   \title{Incompatibility of a comoving Ly$\alpha$ forest with\\
supernova-Ia luminosity distances}
   \author {Jens Thomas \and Hartmut Schulz}
\institute{Astronomisches Institut der Ruhr-Universit\"at, D-44780 Bochum, Germany\\
jthomas, hschulz@astro.ruhr-uni-bochum.de \\} 
\date{Received:FILL IN; accepted:FILL IN}
\offprints{H. Schulz}
   \maketitle
\markboth{J. Thomas \& H. Schulz: Comoving Ly$\alpha$-forest and SNIa data}
{J. Thomas \& H. Schulz: Comoving Ly$\alpha$-forest and SNIa data}
\begin{abstract}
Recently Perlmutter et al. (1999) suggested a positive value of Einstein's 
cosmological constant $\Lambda$ on the basis of luminosity distances from
type-Ia supernovae (the ``SN-method''). However, $\Lambda$ world models 
had earlier been proposed by Hoell \& Priester (1991) and Liebscher et al. (1992a,b) 
on the basis of quasar absorption-line 
data (the ``Q-method''). Employing more general repulsive fluids (``dark energy'')
encompassing the $\Lambda$ component
we quantitatively compare both approaches with each other. 

Fitting the SN-data by a minimum-component model consisting of dark energy + dust (pressureless matter)
yields a closed universe with a large amount of dust exceeding the baryonic content constrained
by big-bang nucleosynthesis. The nature of
the dark energy is hardly constrained. Only when enforcing a flat universe there is a clear tendency
to a dark-energy $\Lambda$ fluid and the `canonical' value $\Omega_{\rm M}\approx 0.3$ for dust.

Conversely, a minimum-component Q-method fit yields a sharply defined, slightly closed model with a 
low dust density ruling out significant pressureless dark matter. The dark-energy component obtains
an equation-of-state ${\cal P}=-0.96 \epsilon$ close to that of a $\Lambda$-fluid (${\cal P}=- \epsilon$).
$\Omega_{\rm M} =0.3$ or a precisely flat spatial geometry are inconsistent with minimum-component models.

It is found that quasar and supernova data sets 
cannot be reconciled with each other via (repulsive ideal fluid+incoherent matter+radiation)-world models.
Compatibility could be reached by drastic expansion of the parameter space
with at least two exotic fluids added to dust and radiation as world constituents. 
If considering such solutions as far-fetched one has to conclude that the Q-method
and the SN-Ia constraints are incompatible. 
\end{abstract}
%
\section{Introduction}
In Newtonian gravitational theory, confined to positive masses as active and passive sources, gravitation is
an exclusively attractive type of interaction. 
In general relativity Einstein (1917) showed by introduction of the cosmological constant $\Lambda$ that gravitation
may be repulsive as well (if $\Lambda > 0$). Homogeneous and isotropic general-relativistic
world models {\em including} the $\Lambda$-term
have been reconsidered in recent years for a variety of reasons.

In the 1990's a tightly constrained world model including
a $\Lambda$ term (the so-called Bonn-Potsdam model; henceforth BP-model) was derived by fitting
quasar absorption lines (Hoell \& Priester 1991a,b; Hoell et al. 1994;
Liebscher 1994; Liebscher et al. 1992a,b).
The BP-model was based on the assumption that Ly$\alpha$-forest quasar absorption
lines situated between the Ly$\alpha$ and Ly$\beta$ emission lines
reproduce the redshifts of comoving galaxy-populated shells (called
`bubble walls' by the authors) in the universe.
The authors considered as virtues that their model
does not require dark matter (except the $\Lambda$-fluid) and has a long age ($3\,10^{10}$ years) providing ample time 
for galaxy formation. If fixing the matter content by the baryons predicted through big-bang nucleosynthesis, it leads
to a large Hubble constant $H_0$. Large values of $H_0$ were claimed in the 
early days of the HST key project on the distance scale. 

During the last few years also other groups went on the `$\Lambda$ band waggon'. One reason is the apparent need
for a nearly flat ($\Omega \approx 1$) universe inferred from cosmic microwave background observations
(Lineweaver 1998, de Bernardis et al. 2000). 
Since the $\Lambda$ term in Einstein's field
equation is equivalent to the presence of a fluid with an energy density constant in space and
time it can be used to fill the universe with `missing energy'. Nucleosynthesis constraints limit the amount
of baryonic matter far below one in $\Omega$ units 
and peculiar-motion observations suggest that also the attractive dark matter falls short to make the universe
flat (Zehavi \& Dekel 1999). 

Recently, the repulsive effect of  $\Lambda>0$ may have been detected more directly.
Observations of supernovae of type Ia, taken as cosmic
`standard candles', appear to require the presence of an accelerating component in the expanding 
universe (Riess et al.\ 1998; Perlmutter et al.\ 1999). 

In view of this added evidence for $\Lambda$, one may ask whether the seemingly pioneering
Q-approach can be reconciled with the new data. However, since the original BP-model significantly 
differs from published fits of $\Lambda$ models to the SN 
data (Perlmutter et al. 1999; from now on we use
the term `SN models' for all models that fit the SN data) we employ a more general approach, which
for our fits is nevertheless based on the same data sets as published BP and SN models.
Rather than only taking $\Lambda$ as representing a repulsive component, we allow for
fluids which have in common
that the equation of state of each fluid yields proportionality between energy density $\epsilon$ and
pressure $\cal P$. This form of the equation of state includes the classical cosmological
matter components dust, radiation and the $\Lambda$ fluid, but also encompasses more general repulsive and
attractive components. 
\section{Theoretical background}
\label{basic}
\subsection{World models}
We here briefly describe the framework of the multifluid world models, which will be fit
to the SN and quasar data. Further details, in particular on their classification, are given in
Thomas \& Schulz (2001). 

We assume the Cosmological Principle, i.e. homogeneity and isotropy of the universe,
leading to the Robertson-Walker (RW) line element. 
Inserting this into Einstein's field equation 
leads to the Friedmann equations
\begin{equation}
\label{fried1}
\frac {\ddot{a}} {a}  =  - \frac {4 \pi G} {3} (\epsilon + 3 {\cal P})
\end{equation} 
and
\begin{equation}
\label{fried2}
H^2(t) \equiv \left( \frac {\dot{a}} {a} \right)^2 = \frac {8\pi G} {3} \epsilon - \frac {k}{a^2}
\end{equation}
$H(t)$ is the Hubble parameter at time $t$, the vacuum velocity of light is set $c=1$. 
$\epsilon = \sum \epsilon_i$ denotes the total energy density of the universe, which
we consider as the added contributions of $N$ ideal fluid components, each with an
energy density $\epsilon_i$. The energy density of each fluid is correlated to its
pressure ${\cal P}$ via an equation of state ${\cal P}=\cal P(\epsilon)$, for which we assume
\begin{equation}
\label{eqs}
{\cal P} = (\gamma - 1) \epsilon.
\end{equation}
The constants $\{ \gamma_i \}$ specify different fluid components. This simple form 
encompasses 
the well known cosmic fluids like (e.g. Kolb \& Turner 1990)
\begin{itemize}
\item dust (${\cal P}=0; \gamma=1$),
\item radiation and any other extremely relativistic matter (${\cal P}=\epsilon/3; 
\gamma=4/3$), 
\item the constant-density $\Lambda$-fluid (${\cal P}=-\epsilon; \gamma=0$), 
\item a network of slow cosmic strings (${\cal P}=-\epsilon/3; \gamma=2/3$) or 
\item domain walls (${\cal P}=-(2/3)\epsilon; \gamma=1/3$). 
\end{itemize}
The term ``dust'' should not be confused with the
solid-particle component of the interstellar medium as is common
in astronomy. We here follow the practice of relativists
to designate non-relativistic (${\cal P}=0$) incoherent matter as dust.
In particular, dust includes all baryonic matter and
cold dark matter in its simplest form.

Apart from dust and radiation the physical 
background of cosmic fluids with equation of state\,(\ref{eqs}) is hardly understood.
Therefore, in case of $\gamma \ne 1$ or $4/3$ we use the designation {\em exotic fluid}.

We confine the parameter of the equation of state to the range
\begin{equation}
\label{gam}
0  \leq  \gamma \leq 2 \ \ \mbox{implying} \ \ c_s \leq c
\end{equation}
where the speed of sound $c_s$ is restricted not to surpass the 
velocity of light $c$.

In standard world models, positive as well as
negative values of the cosmological constant $\Lambda$ have been considered. Since in the
fluid description $\Lambda$ corresponds to a constant density, it is suggestive to allow for positive {\em and}
negative energy densities of other $\gamma$ fluids in our phenomenological approach as 
well. 

After introducing the normalized density $\Omega_i$, the normalized scale factor $x(t)$ and the parameter of state
$\alpha_i$ by the equations\footnote{By $H_0, \ a_0, \ {\rm and} \ \epsilon_{i0}$ we denote the values of
$H(t)$, $a(t)$ and $\epsilon_i = \epsilon_i(t)$ at the present epoch $t=t_0$.}
$$
\Omega_i \equiv \frac {8\pi G} {3 H_0^2} \epsilon_{0i}, \ \ x(t) \equiv \frac{a(t)}{a_0} \ \ {\rm and} \ \ 
\alpha_i \equiv 2-3\gamma_i
$$
Eqs.\,(\ref{fried1}), (\ref{fried2}) obtain the more suitable  form 
\begin{equation}
\label{exp1}
\ddot{x} = \frac {H_0^2} {2 x} \sum_i \alpha_i \Omega_i x^{\alpha_i} 
\end{equation}
\begin{equation}
\label{exp2}
\dot{x}^2 = H_0^2 \left(\sum_i \Omega_i x^{\alpha_i} +1 - \sum_i \Omega_i \right) 
\end{equation}
To arrive at Eq.\,(\ref{exp2}) one has to use the present-epoch ($t=t_0$) version of Eq.\,(\ref{fried2}) to eliminate the
curvature index $k$ 
\begin{equation}
\label{curv}
k = a_0^2 H_0^2 \left( \sum_i \Omega_i - 1 \right).
\end{equation}

Henceforth we call components with $\alpha > 0$ {\em $\Lambda$-like}, those with $\alpha < 0$ 
{\em dustlike} because of their dynamical {\em similarity} to the $\Lambda$-fluid ($\alpha = 2$) and 
cosmic dust ($\alpha = -1$), respectively.

According to Eq.\,(\ref{exp1}), fluids obeying $\Omega \alpha > 0$ accelerate the cosmic expansion, we call 
them {\em repulsive}. Analogously, {\em attractive} fluids are characterized by $\Omega \alpha <0$.
\subsection{Method to fit the SN-Ia luminosity distances}
\label{snmet}
We here assume that the corrections for extinction by intervening material, waveband width and shift (K correction)
as well as peak magnitude via the shape of light-curve have been appropriately applied so that
the SNe-Ia can be utilized as {\em relative} standard candles as discussed in Perlmutter et al. (1997, 1999). 
In a geometrical cosmological model the result of light propagation between an observer at $z=0$ receiving
light from an emitter at redshift $z$ is subsumed under the term `luminosity distance', which can be used like
an Euclidean distance that leads to inverse-square light dilution. For a given model,
parameterized by a set of fluid parameters $\{(\Omega_i, \alpha_i)\}$ luminosity distance or the
corresponding distance module are functions of $z$. Since the absolute peak magnitude of the SNe is 
assumed to be known to a constant, 
one can compare the observed variation of apparent peak magnitude with $z$
with the shape of theoretical distance-module curves given by
\begin{eqnarray}
\label{sn}
m &=& C + 5 \, \log \left(
\frac{1+z}{\sqrt{\left| \sum_i \Omega_i - 1 \right| }} \, 
\left\{
\begin{array}{l@{...}l}
\sin(f) & \ k=1\\
f & \ k=0 \\
\sinh(f) & \ k=-1 \\
\end{array}
\right\} \right) \nonumber \\
f &=& f(z) =  
\int\limits_{0}^{z}
\frac{(1+z')^{-1}\sqrt{\left| \sum_i \Omega_i - 1 \right|} \, \der z'}
{\sqrt{\sum_i \Omega_i \left( \left( 1+z' \right)^{-\alpha_i} -1 \right)+ 1}}
.  
\end{eqnarray} 
Here $m$ is the corrected apparent magnitude and $C$ is a constant.

The supernova data set is
given in  Perlmutter et al. (1999). It contains redshift and magnitude data of 42 high redshift SN ($0.172 < z < 0.830$) and additionally 
of 18 low redshift SN ($0.014 < z < 0.101$) which are needed to calibrate the method (cf. Perlmutter et al. 1997).

We adopted this relatively homogeneous set as representative 
for a demonstration of the SN-method. Graphs and results of the
measurements by Riess et al. (1998, 1999) show these to be pleasingly
similar. Further SN luminosity distances are rapidly accumulating and
will provide stronger constraints in the future.
However, it is not our goal to look for the `ultimate' fitting model.
Rather we want to obtain a typical range of fitting models that
include our more general fluids.

In order to obtain best-fit values for the fluid parameters we minimized the function
\begin{equation}
\label{fitsn}
\chi^2 = \sum_{j=1}^{60} \frac{\left( m(z_j,\Omega_1, \ldots, \Omega_N,\alpha_1, \ldots,\alpha_N) - m_j \right)^2}{\sigma_j^2}.
\end{equation}

In contrast to the Q-method described below,
we note that the apparent magnitude of the SNe
only contains information about the {\em integrated} effect
of cosmic evolution since the SN explosion has happened. Due to this smearing-out effect
the evaluation of parameters is not tightly constrained.
\subsection{Q-method to fit the Ly$\alpha$-forest}
\label{qmet}
The basic idea underlying the Q-method is that the Ly-$\alpha$-forest is caused by
intersections of the quasar light with the boundaries of an inhomogeneous bubble structure of the (baryonic)
matter (Hoell \& Priester 1991a,b; Hoell et al. 1994;
Liebscher 1994; Liebscher et al. 1992a,b). The differences $\Delta z$ between the 
redshifts $z$ of two absorption-lines are observed to be small, so that they are taken as
infinitesimal. Then $\Delta z$ can be directly converted into the model-dependent 
cosmic expansion rate $H(z)$ at $z$, the epoch of the twin absorption, if the distance
between the absorbing bubble walls (the void diameter) is known. The BP group
{\em assumed} a {\em constant} void diameter $R_{\rm B}$ in the {\em comoving} frame.
With these presumptions (and using the abbreviation $y \equiv 1+z$), one finds in close analogy 
to the Q-procedure (as described, e.g., in Liebscher et al. 1992a)
\begin{equation}
\label{pr}
{\Delta z}^2 = R_{\rm B}^2 a_0^2 H_0^2 \left( \sum_i \Omega_i y^{(2-\alpha_i)} + \left(1 - \sum_i \Omega_i \right) y^2 \right).
\end{equation}
In the present case we minimize the function
\begin{equation}
\label{chipr}
\chi^2 = \sum_{j=1}^{14} \frac{\left( {\Delta z}^2(z_j,\Omega_1, \ldots, \Omega_N,\alpha_1, \ldots,\alpha_N) 
- {\Delta z_j}^2  \right)^2}{\sigma_j^2}.
\end{equation}

The data set we used is given in Liebscher et al. (1992a). There are 12 absorption-line differences derived
from the spectra of four quasars with $2.25 < z < 4.35$ together with the data of two nearby voids at $z=0.08$ and
$z=0.03$, respectively.

In another work (Liebscher et al. 1992b) the BP-group published further absorption line data, however the cosmic
parameters obtained from the enlarged data sample remain nearly the same.

The advent of space-based and ground-based
high-resolution spectrographs led to Lyman forest data at lower $z$
than available for the BP-group (Weymann et al. 1998, Christiani et al.
2000, Kim et al. 2001). However, we have refrained to extract 
mean $\Delta z$ values from these measurements, leaving this tedious task
to defenders of the Q-method. It is shown below that the data set
employed here leads to by far sufficient constraints for the Q-method
to unveil strong differences with the SN-method.
\section{Results of the fits}
\subsection{SN-Ia data}
For the derivation of cosmic parameters we considered various cosmic fluids 
given by $(\Omega, \alpha)$ or $\{\Omega_i, \alpha_i \}$ in addition to
incoherent dust with density $\Omega_{\rm M}$ (the background radiation component 
could be usually neglected because of the low redshifts). The $\chi^2$ values are not
normalized and only meaningful within the same group of fits.
\paragraph{Case 1: Dust + exotic component.} With only {\em one} arbitrary exotic fluid the best fit yields
($\nu$ denotes the degrees of freedom of the fit) 
\begin{eqnarray}
\Omega_{\rm M} &=& 0.86\ \pm \ 0.58 \nonumber \\
\Omega &=& 1.38\ \pm \ 0.73 \nonumber \\
\alpha &=& 2\ \pm1.9 \nonumber \\
\chi^2 &=& 102 \ , \ \nu=58\nonumber .
\end{eqnarray}
Interpreting the resulting $\alpha \approx 2$ as pointing towards $\Lambda$ as dark energy component
is premature. The amount of uncertainty in $\alpha$
suggests that nearly any $\Lambda$-like fluid can be fit to the data. For instance, when fixing $\alpha=1$
we found no difference in $\chi^2$ :
\begin{eqnarray}
\Omega_{\rm M} &=& 0.65\ \pm \ 0.41 \nonumber \\
\Omega &=& 2.41\ \pm \ 1.11 \nonumber \\
\chi^2 &=& 102 \ , \ \nu=58\nonumber. 
\end{eqnarray}

With increasing $\alpha$ the age of the universe in the best-fit models decreases 
($\alpha = 1 \to < 20 \, {\rm Gyr}, \alpha = 2 \to < 15 \, {\rm Gyr}$) 
and concomittingly $\Omega_{\rm M}$ decreases.

\paragraph{Case 2: Dust fixed+free exotic component.} The dust densities derived from the above fits are high 
compared to values usually derived with other methods (e.g. Zehavi \& Dekel 1999).
If one assumes lower dust densities, $\alpha$ decreases (albeit remaining positive) thereby weakening 
the tendency to $\Lambda$ as repulsive fluid. For $\Omega_{\rm M} = 0.3$ (a typical result from
observations of peculiar velocities) we found $\alpha = 0.6$, 
for $\Omega_{\rm M} = 0.03$ (closer to nucleosynthesis constraints for baryons) the result was
$\alpha = 0.5$ (cf. Tab.~\ref{zssnfit}).

\begin{figure}[ht]
\resizebox{\hsize}{!}{\includegraphics[angle=-90]{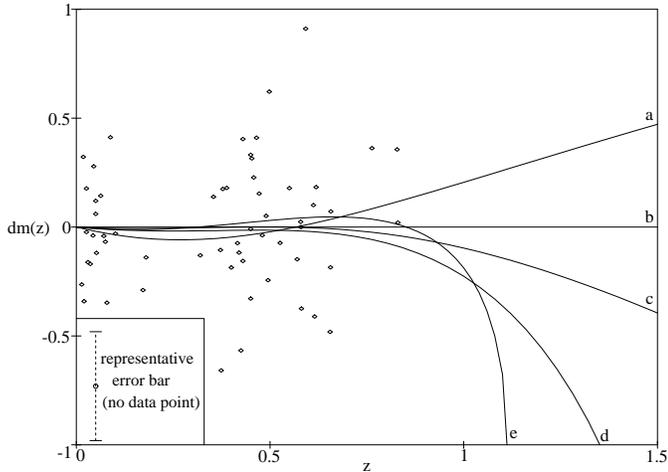}}
\caption{SN world models of Tab.~\ref{zssnfit} relative to the best fit model (horizontal line). 
$dm(z) \equiv m_{\rm model}(z) - m_{\rm bestfit}(z)$. A typical error bar is shown in the insert (lower left). For 
individual errors of the data see Perlmutter et al. 1999.}
\label{snfits}
\end{figure}

\paragraph{Case 3: Flat models.} With no restriction for $\Omega_{\rm M}$ but considering only 
flat universes the $\Lambda$-fluid turns out to be the best-fit exotic fluid:
\begin{eqnarray}
\Omega_{\rm M} &=& 0.30\ \pm \ 0.04\nonumber \\
\alpha &=& 2\ \pm \ 0.04 \nonumber \\
\chi^2 &=& 104 \ , \ \nu=58\nonumber.
\end{eqnarray}
Now, in contrast to case 1  the minimum of $\chi^2$ is rather well defined and, statistically,  there is no 
likely alternative to a dark-energy component close to the $\Lambda$ fluid. 
We nevertheless emphasize that - when relaxing the assumption of flatness - closed world models
lead to slightly better (lower $\chi^2$) fits (compare Fig.~\ref{snfits}).

Among the fluid parameter sets leading to world models compatible with the observations, we found
models without a big bang (however, all best fit models start with a big bang). 
If we want to transfer the cosmological standard model of the formation of light elements 
to non-big-bang universes, we have to require an epoch characterized by physical conditions similar to 
those in the nucleosynthesis epoch of a successful big-bang model.
One of the necessary conditions would be a temperature exceeding $10^9$ K, which forces the minimum of the scale
factor to fall short of $x_{\rm min} < 10^{-9}$. This condition rules out successfully fitting non-big-bang models. 

However, the situation changes when allowing for more than one exotic fluid. 
For instance, we found flat models with
two exotic fluids (plus dust) that fit the data and reach a minimum $x_{\rm min} < 10^{-9}$.
A particular example is given by the following parameters:
\begin{eqnarray}
\Omega_{\rm M} &=& 0.3 \nonumber \\
\Omega_{\rm \gamma} &=& 5.9 \, 10^{-5} \nonumber \\
\Omega_1 &=& -5.90003 \, 10^{-14} \ , \ \alpha_1 = -3 \nonumber \\ 
\Omega_2 &=& 1 - \Omega_{\rm M} - \Omega_{\rm \gamma} - \Omega_1 \ , \ \alpha_2 = 1.5 \nonumber \\
\chi^2 &=& 102 \ , \ \nu=58\nonumber. 
\end{eqnarray}
In this model we have also included the component of cosmic background radiation $\Omega_{\rm \gamma}$
because of the desired high temperatures and the correspondingly small $x$ (high $z$) at which
the dynamical effect of the radiation component (or any classical relativistic component)
is strongly enhanced although negligible today. 
\begin{table}[ht]
\caption{Results of the supernova-fits. $\chi^2_{\rm red} \equiv \chi^2 / \nu$ denotes the reduced $\chi^2$ with 
$\nu$ degrees of freedom. For the models below $\nu = 58$.}
\label{zssnfit}
\begin{center}
\begin{tabular}{l|l|l|c}
Model & Parameter & $\chi^2_{\rm red}$ & in Fig.~\ref{snfits}\\
\hline
flat & $\Omega_{\rm M} = 0.30$ & 1.79 & a \\
         & $\Omega = 0.70 \ , \ \alpha =2$ & & \\
        & $\sum \Omega_i=1$ fixed & & \\ 
$\Lambda$-fluid & $\Omega_{\rm M} = 0.86$ & 1.76 & b \\
                          & $\Omega = 1.38 \ , \ \alpha=2$ & & \\
                          & $\alpha=2$ fixed & & \\
Domain walls   & $\Omega_{\rm M} = 0.65$ & 1.76 & c \\
                          & $\Omega = 2.41 \ , \ \alpha = 1$ & & \\
                         & $\alpha=1$ fixed & & \\
matter + CDM&$\Omega_{\rm M} = 0.3$ & 1.76 & d \\
                           &$\Omega = 3.2 \ , \ \alpha=0.6$ & & \\
                        &$\Omega_{\rm M} = 0.3$ fixed & & \\
matter&$\Omega_{\rm M} =0.03$& 1.76 & e \\
                           &$\Omega=3.3 \ , \ \alpha=0.5$ & & \\
                         &$\Omega_{\rm M} = 0.03$ fixed & & 
\end{tabular}
\end{center}
\end{table}
%
%
%
\subsection{Quasar data}
Fitting the quasar data with a model consisting of dust and only one arbitrary exotic fluid we obtain
\begin{eqnarray}
\Omega_{\rm M} &=& 0.0162 \ \pm \ 0.0003 \nonumber \\
\Omega &=& 1.100 \ \pm \ 0.001 \nonumber \\
\alpha &=& 1.87 \ \pm \ 0.01 \nonumber \\
\chi^2 &=& 10  \ , \ \nu=11\nonumber.
\end{eqnarray}

This model favours a low dust content and a repulsive fluid with properties close to that
of a $\Lambda$-fluid. When attributing the dust component exclusively
to the baryon content predicted by big-bang nucleosynthesis it requires a large $H_0$. In comparison to the SN fits it 
appears most striking that the minimum of $\chi^2$ is sharp 
and other $\Lambda$-like fluids are practically ruled out.

We attempted to fit models including cold dark matter, e.g. by fixing $\Omega_{\rm M} = 0.3$, but found
no match with the data ($\chi^2 > 100\,000$). This statement no longer holds if more than one exotic
fluid is invoked. In this more general case we found models when fixing $\Omega_{\rm M} = 0.3$, for parameters
of state of the exotic fluids in the range $\alpha_{1/2} < -1$ and $\alpha_{1/2} > 0$. One example is the model
\begin{eqnarray}
\Omega_{\rm M} &=& 0.3 \nonumber \\
\Omega_1 &=& -0.060 \ , \ \alpha_1 = -1.56 \nonumber \\
\Omega_2 &=& 1.879 \ , \ \alpha_2 = 1 \nonumber \\
\chi^2 &=& 11 \ , \ \nu=11\nonumber.
\end{eqnarray}

All best-fit models with one exotic fluid are nearly flat, whereas {\em exactly flat} (within
the parameter errors) models were not found to fit to the data. Exactly flat models with {\em two exotic fluids} exist. 
Thereby for each of the exotic fluids the parameter of state must obey $\alpha_{1/2} \geq -0.58$. 
An example is given by the following parameters:
\begin{eqnarray}
\Omega_{\rm M} &=& 0.037 \nonumber \\
\Omega_1 &=& -0.082 \ , \ \alpha_1 = -0.58 \nonumber \\
\Omega_2 &=& 1 - \Omega_{\rm M} - \Omega_1 \ , \ \alpha_2 = 2 \nonumber \\
\chi^2 &=& 11 \ , \ \nu=11\nonumber.
\end{eqnarray}

The ages of the universe in those fitting models that start with a big bang is high in comparison
to those of SN-world models
($\sim 35 \, {\rm Gyr}$ for $H_0 = 65 \, \mathrm{km/s/Mpc}$). For successful models with an exotic fluid and no big bang the same holds as for the
corresponding SN-models:  they do not contain an epoch with $x_{\rm min} < 10^{-9}$ ruling out the
possibility of light-element nucleosynthesis in the same way as in standard-big-bang universes.
\begin{table}
\caption{Results of the Quasar-Fits.}
\label{result}
\begin{center}
\begin{tabular}{l|l|l|c}
Model & Parameter & $\chi^2_{\rm red}$ & in Fig.~\ref{prfits}\\
\hline
$\Lambda$-fluid & $\Omega_{\rm M} = 0.0122 \pm 0.0003$ & 0.92 & D \\
         & $\Omega = 1.073 \pm 0.001 \ , \ \alpha =2$ & & \\
        & $\alpha=2$ fixed & & \\ 
best fit & $\Omega_{\rm M} = 0.0162 \pm 0.0003$ & 0.91 & B \\
                          & $\Omega = 1.100 \pm 0.001$ & & \\
                          &  $\alpha=1.87 \pm 0.01$ & & \\
matter   & $\Omega_{\rm M} = 0.0323 \pm 0.0003$ & 1.17 & C \\
                          & $\Omega = 1.220 \pm 0.001 \ , \ \alpha = 1.5$ & & \\
                         & $\alpha=1.5$ fixed & & \\
flat &$\Omega_{\rm M} = 0.037$ & 1.00 & A \\
(example)   &$\Omega_1 = -0.082 \ , \ \alpha_1=-0.58$ & & \\
                   &$\Omega_2 = 1.045 \ , \ \alpha_2 = 2$ & & \\
                        &$\sum \Omega_i = 1$ and $\alpha_2 = 2$ fixed& & \\
matter  &$\Omega_{\rm M} =0.3$& 1.00 &  - \\
+ CDM       &$\Omega_1=-0.060 \ , \ \alpha_1=-1.56$ & & \\
(example)      & $\Omega_2=1.879 \ , \ \alpha_2=1$ & & \\
                         &$\Omega_{\rm M} = 0.3$ and $\alpha_2 = 1$ fixed & & 
\end{tabular}
\end{center}
\end{table}
\begin{figure}[ht]
\resizebox{\hsize}{!}{\includegraphics[angle=-90]{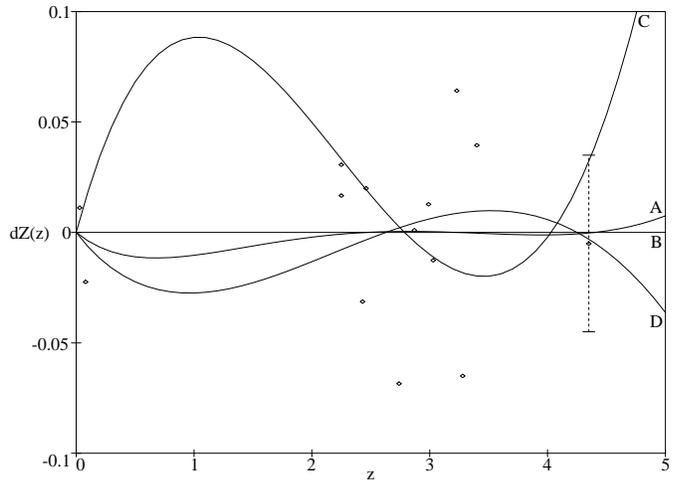}}
\caption{Q-world models of Tab.~\ref{result} relative to the best fit model B ($\alpha = 1.87$). 
$dZ(z) \equiv {\Delta z}_{\rm model}^2(z) - {\Delta z}_{\rm bestfit}^2(z)$. Only one error bar, representative for
all data points, is shown.}
\label{prfits}
\end{figure}
\subsection{Comparison of the results}
\label{compofres}
For the case of world models consisting of dust and one exotic fluid both methods
agree in the requirement of a {\em repulsive} fluid leading to an {\em accelerated} universe
even for low $z$. Within the framework
of generalized ideal fluids presented in Thomas \& Schulz (2001) such a repulsive fluid could 
be either $\Lambda$-like with positive energy density or dustlike with negative density (cf. Sect.\,2.1). 
Both the SN luminosity distance and the quasar absorption-line method lead 
to a $\Lambda$-like exotic repulsive fluid. In both cases, the data themselves can be fit
sufficiently well with models containing only one exotic component.

However, as Fig.~\ref{snfits} shows, the SN-method does not allow to decide which equation of state describes
the exotic fluid. A large part of the degeneracy in the exotic component is caused by the relatively low redshifts
so far reached by SN observations. Therefore at first glance,  the Q-method, reaching
to comparatively high redshifts close to $z \approx 4.5$,  appears to be superior to narrow down the exotic component.
Caution has to be applied to this suggestive conclusion because the fits of both methods differ significantly.

Typically, the SN data lead to closed models with
large dust densities $\Omega_{\rm M}$ (allowing for cold dark matter).
Despite presence of a repulsive component these models yield a relatively low age of the 
universe - roughly between 10 and 20 Gyr 
($H_0 = 65 \, \mathrm{km/s/Mpc}$). 
As pointed out above, the kind of accelerating component is hardly constrained: nearly any 
$\Lambda$-like component can be tuned to
match the observations. Only when a flat geometry is required this degeneracy breaks down and the
SNe Ia point towards $\Lambda$ as dark-energy fluid.

On the other hand, the quasar models usually turn out to be nearly flat, require low dust densities,
which exclude substantial amounts of cold dark matter. They lead 
to values of the age of the universe of at least 30 Gyr ($H_0$ as above). Furthermore,
there is a strong tendency to $\Lambda$ as repulsive component rather than something `more exotic'.

\begin{figure}[ht]
\resizebox{\hsize}{!}{\includegraphics[angle=-90]{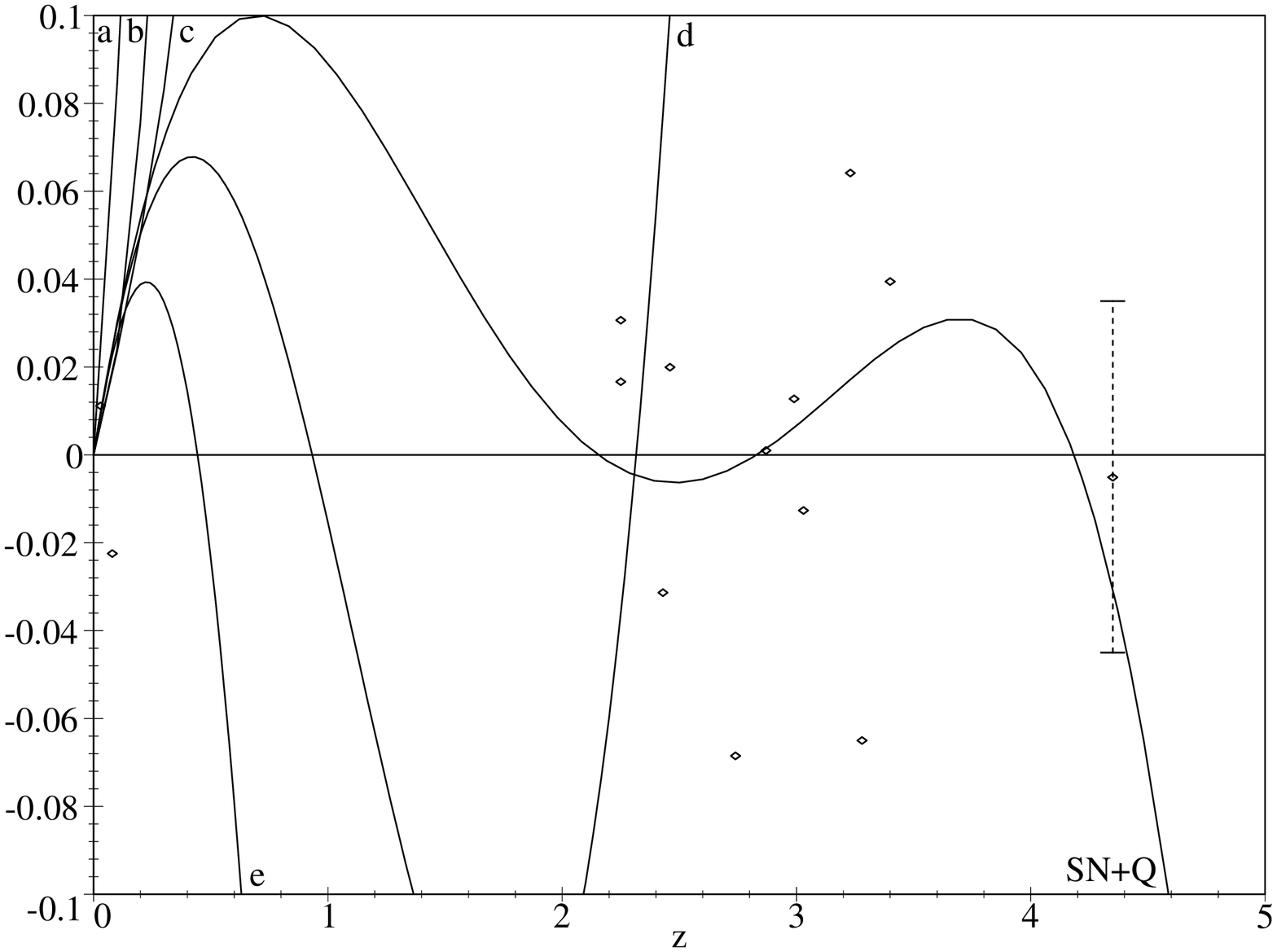}}
\resizebox{\hsize}{!}{\includegraphics[angle=-90]{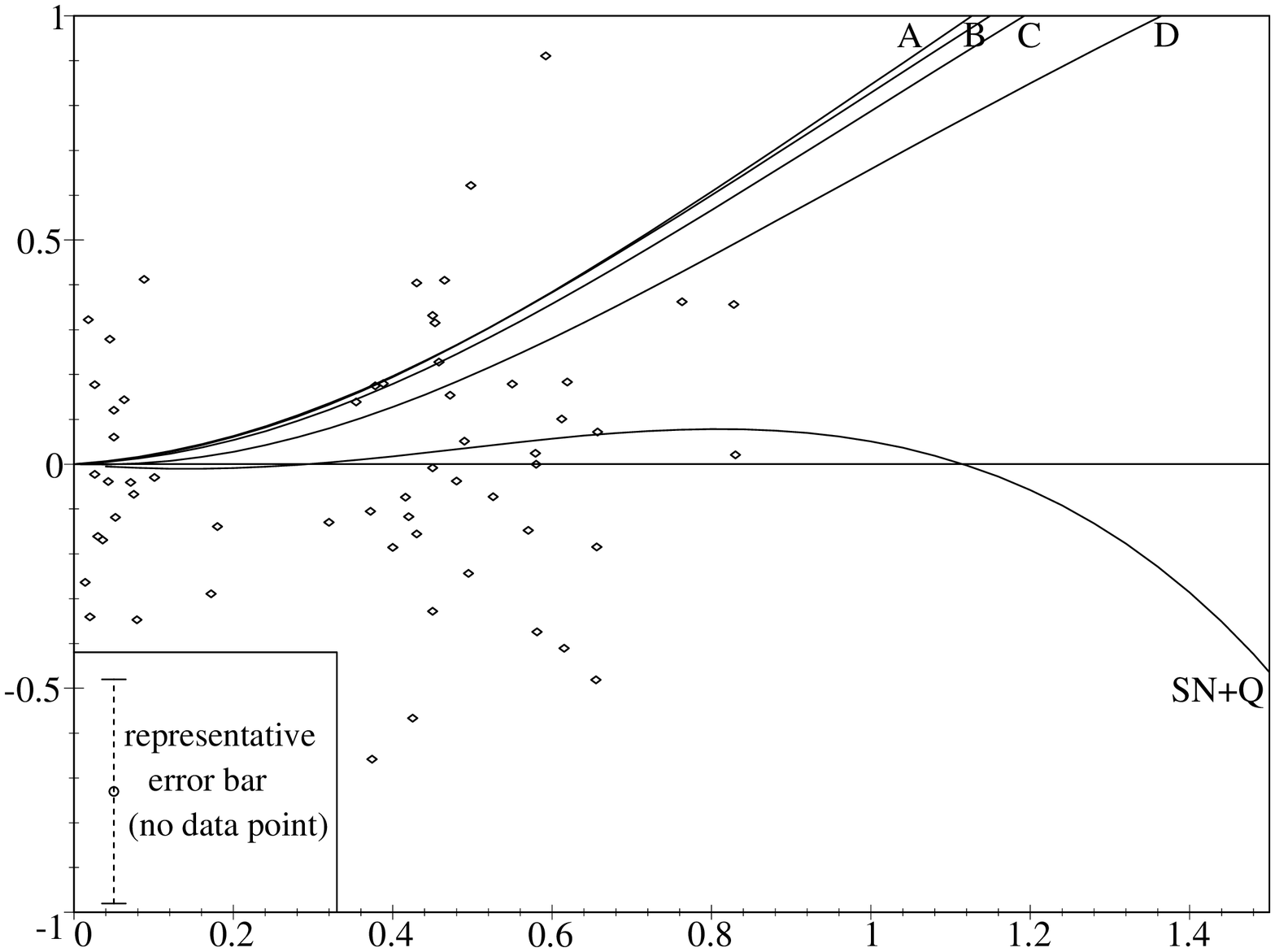}}
\caption{The upper diagram shows the predicted absorption-line differences of the SN world models of 
Tab.~\ref{zssnfit} (cf. Fig.\,2). In the lower diagram the differences of the 
supernova luminosities (relative to the best fit; cf. Fig.\,1) predicted by the Q-models of Tab.~\ref{result} are drawn 
together with the observed data. The curves ``SN+Q'' represent the model of Sect.~\ref{compofres} that fits to 
both data sets.}
\label{slfits}
\end{figure}

Actually, the SN models are not compatible with the Q-method and vice versa. For illustration the upper 
part of Fig.~\ref{slfits} shows the absorption-line redshift
differences predicted by the SN models, the lower part shows the SN-luminosity residuals
predicted by the Q-models.
In order to fulfil the conditions set by both methods with one model, one is forced to consider at least two 
repulsive fluids. For example, the following model ``SN+Q'', which is also shown in Fig.~\ref{slfits}, fits both 
data sets:
\begin{eqnarray}
\Omega_{\rm M} &=& 0.5 \nonumber \\
\Omega_1 &=& 2.44 \ , \ \alpha_1 = 0.8 \nonumber \\
\Omega_2 &=& -0.12 \ , \ \alpha_2 =-1.5 \nonumber \\
\chi^2_{\rm SN} &=& 102\nonumber \\
\chi^2_{\rm Q} &=& 11.\nonumber
\end{eqnarray}

\begin{figure}[ht]
\begin{center}
\resizebox{\hsize}{!}{\includegraphics[angle=-90]{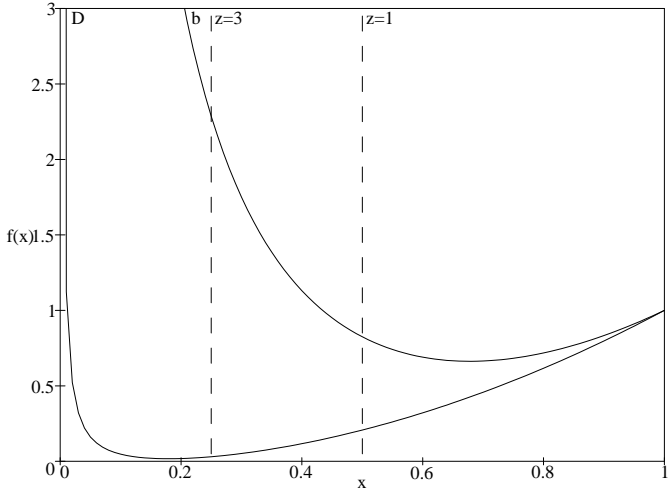}}
\caption{$f(x) \equiv \dot{x}^2(x)$ for the Q-model ``D'' (cf. Tab.~\ref{result}) and the SN-model ``b'' (cf. Tab.~\ref{zssnfit}). 
Supernova data points are mainly located within the range $0.5 < x < 1$, the employed 
quasar Ly-forest lines arise between $x=0.2$ and $x \approx 0.35$.}
\label{expfit}
\end{center}
\end{figure}

The kinematical differences of the SN-models and the Q-models are illustrated in Fig.\, \ref{expfit}, for which we
have chosen the model ``b'' of Tab.\, \ref{zssnfit} and ``D'' of Tab.\, \ref{result} as representative examples for
SN-models and Q-models, respectively. The figure shows $f(x) \equiv \dot{x}^2(x)$, e.g. the expansion rate of the
models as a function of the scale factor $x$.
In the range of the absorption line data ($0.2 < x < 0.35$) the Q-models expand slowly,
as compared to the SN-models, leading to relatively small absorption line differences for
given sizes of comoving bubbles. In most SN-models the same comoving
bubbles would produce much larger differences of the lines (cf. Fig.\, 3, upper panel). 
SN models cannot fit a slow observed evolution of bubble-wall differences because
the development of brightness of the 
supernovae for $0.5 < x < 1$ with redshift $z$ forces a rapid expansion. In the Q-models the supernovae would
simply be too dark.

The nearly static phase at $x \approx 0.2$ strongly contributes to the high ages of the Q-models.
\subsection{Remark on void evolution}
We attempted to reconcile the SN world models with a bubble-wall
Ly$\alpha$-forest via assuming various void evolution models.
To this end we replaced the constant void-parameter
$R_{\rm B}$ in the Q-method by an ansatz $R_{\rm B}(z) = a_1 \, z^{b_1}+ a_2 \, z^{b_2}$.
However, this approach yielded bad fits with at least ten times
larger chisquares than the Q-fits.

This result does not appear surprising in view of the fact that
a comoving Ly$\alpha$-forest requires an extended nearly static epoch in the
Q-fits (e.g. see Fig.~\ref{expfit}). To force agreement with the
rapidly expanding SN models cannot even be achieved by our
double-power-law ansatz.
\section{Discussion}
Unless we are invoking complicated multi-parameter models, the Q-models and the SN fits with
world models composed out of $\cal P \propto \epsilon$ - fluids have been shown to be incompatible.
One possibility would be that Robertson-Walker models themselves are inadequate approximations
to an inhomogeneous universe. However, this unsolved problem reviewed by Krasinski (1997) 
is beyond the scope of the present work.

Taking for granted the standard view that RW models are fair approximations of the global
behavior of the universe we confine the discussion to possible pitfalls inherent to the Q-approach
or to the SN fitting procedure.
\subsection{The Q-approach}
\label{qres}
The Q-approach rests on at least two assumptions: (i) voids and their `bubble walls' are stable 
structures, whose comoving average size is constant  from $z=0$ to at least $z=4.4$;
(ii) the locations of the bubble walls are uniquely determined by the quasar absorption lines selected
by the Bonn-Potsdam group. 

(i) During its early linear phase, 
the size of a small cosmic perturbation will be constant in a comoving frame
(Padmanabhan 1993). However, the comoving size of a void will grow nonlinearly when its density contrast
exceeds an appreciable fraction of unity (Friedmann \& Piran 2000). Since 
current voids have a density contrast of about $\sim (-0.8)$, they have already 
reached the nonlinear stage (El-Ad \& Piran 2000). Insofar it appears to be a critical point
that the Q-method has not yet been applied {\em together} with a self-consistent model of void evolution. 

(ii) In a comprehensive review, Rauch (1996) summarized the complicated history of analysis of the
Ly$\alpha$-forest. A general absence of ``voids"  in the ``typical'' forest (lines of
column density below $10^{14}$ cm$^{-2}$) is claimed by statistical studies. Backed by numerical 
simulations of structure formation,
the Ly-forest lines are considered to originate in an interconnected web of material sheets and
filaments. These simulations utilize appreciable amounts of dark matter, which is at variance with the
low-density baryonic Q-models. At low redshift, there appears to be a trend that the Ly$\alpha$ absorbers
trace the large-scale distribution of galaxies (although absorption systems {\em within} voids were also found). 
However, the Q-approach, as proposed by the BP group, assumes that coincidence of galaxy sheets and 
absorber sheets has persisted since epochs at large redshifts. This contrasts with current belief 
of strong evolution at  relatively low redshift. 

Summarizing, there is disagreement between the BP-group and other workers in this field
about the origin of the Lyman-forest in a cosmological context.
\subsection{SNe-Ia as standard candles}
\label{snres}
In the Perlmutter et al. (1999) sample the evidence for an accelerating component could
vanish if there were a systematic effect making SNe at high redshift dimmer
by $\sim 0.25$ mag relative to those at low $z$. Such a dimming might be 
caused by (i) material between the SNe Ia and the earth. Alternatively (or in addition) 
the possibility of (ii) intrinsic differences between
high-$z$ and low-$z$ SNe has to be considered. (i): Since SNe with obvious reddening are 
not used for deriving luminosity distances Aguirre \& Haiman (2000) discussed the possibility of 
'neutral` intergalactic solid particles making SNe appear fainter without reddening their spectrum.  
It turned out to be a {\em theoretical possibility}, which, if causing the total SNe dimming, should show up in
future measurements of the cosmic far infrared background.  (ii) Systematic differences in the 
spectra of low-$z$ and high-$z$ SNe have not been found.
Comparison of light-curves even showed the widening with $z$ (``time dilation'') due to cosmological expansion 
(Leibundgut et al. 1996). There was concern about significantly shorter rise time of high-$z$ compared
to low-$z$ SNe (Riess et al. 1999), which was, however, narrowed down to $<2\sigma$ {\em agreement}
by Aldering et al. (2000). Despite a controversy on the progenitors and early stages of
their explosion history, the empirically observed homogeneity seems so far to be
fairly well established. Some key tests are nevertheless still to be made, one of which being
a significant extension of the sample to $z \sim 1.3$. A potential shape of the 
magnitude-$z$ curve as for a transition from decelerated to
accelerated expansion could hardly be mimicked by systematic or evolutionary
effects  (Livio 2000).
\section{Summary and conclusions}
We have compared the methods to derive world models from a kind of 
luminosity distance-redshift relation given by SNe Ia observations (the SN approach) 
and, on the other hand, from Ly$\alpha$-forest lines when one assumes that these arise
in galaxy-populated walls and filaments around cosmic voids (the Q-approach). While the original authors essentially
focused on world models with $\Lambda$ plus normal incoherent matter (and radiation, which, however,
only plays a role at dense stages of the universe, e.g. near a big bang) we allowed for a more
general class of cosmic fluids only restricted by an equation of state $\cal P \propto \epsilon$.

However, even within this more general approach world models fit to SNe Ia cannot be reconciled
with those fit to the quasar Ly$\alpha$ redshift distribution\footnote{Only less reasonable 
multi-parameter models can be adjusted to fit both approaches.}. A minimum-component Q-model tends to a
baryonic universe with correspondingly low $\Omega_{\rm M}$  and to $\Lambda$ as repulsive component.
It is therefore incompatible with a number of independent observations that suggest a universe with
dark-matter content $\Omega_{\rm CDM} \sim 0.3$. Structure formation is an open issue in the Q-scenario, although 
the large ages of the Q-models combined with a long period of low expansion may ease galaxy formation.
 
Simple dust+dark energy models fit to the SN-data tend to be closed and require non-baryonic dark matter. 
There is no strong case for $\Lambda$ as dark-energy component unless the models are
forced to be flat.
Generally, there is a wide range of world models fitting the SN-Ia data, which can be 
reconciled with a dark-matter
dominated attractive $\Omega_{\rm M}$, a nearly flat universe, structure formation and current age constraints. 

Despite apparent independent support for the SN-method, it is not our intention to weigh the pros and cons 
of either method. The main conclusion is that there is a deep 
incompatibility between the Q- and the SN method, even if allowing for rather general world models.

This suggests that at least one of the two methods is doubtful or based on
systematic errors requiring critical scrutiny on the underlying 
assumptions (see Sects.~\ref{snmet} and \ref{qmet}). For the Q-method it is mainly the
bold identification of the low-column-density Lyman forest to represent 
a comoving `cell structure' of the galaxy distribution (Sect.~\ref{qres}). For the 
SN-method, although more strongly based on {\em empirical} calibration, there
are still suspicions on some remaining systematic effect (Sect.~\ref{snres}). 

\end{document}